 \documentclass[twocolumn,aps,tightenlines,floats]{revtex4}
\usepackage{graphics}
\usepackage{graphicx}
\usepackage{epsf}
 \usepackage{epsfig}
 \def\bea{\begin{eqnarray}}
 \def\eea{\end{eqnarray}}
 \def\beqs{\begin{eqnarray*}}
 \def\eeqs{\end{eqnarray*}}
 
 \newcommand{\be}{\begin{equation}}
 \newcommand{\ee}{\end{equation}}

\begin{document}

\parbox{1.5in}{ \leftline{JLAB-THY-04-7}
                \leftline{WM-04-101}
                \leftline{}\leftline{}\leftline{}
                \leftline{}}


 \vspace{-1.5in}

 \title
 {\bf Regge Behavior of DIS Structure Functions}

\author
 {Franz Gross$^{1}$, I.V.Musatov$^{1}$, Yu.A.Simonov$^{1,2}$}

\address
{ $^1$Theory Group, Jefferson Lab,  Newport News, VA
23606, USA
 \\
 $^2$ State Research Center, Institute of Theoretical and
Experimental Physics, Moscow, Russia }

\date{\today}

 \begin{abstract}

Building on previous works of the mid 1960's, we construct an
integral equation for forward elastic scattering ($t=0$) at
arbitrary virtuality $Q^2$ and large $s=W^2$.  This equation sums
the ladder production of massless intermediate bosons to all
orders, and the solution exhibits Regge behavior.  The equation is
used to study scattering in a simple $\chi^\dagger\chi\phi$ scalar
theory, where it is solved appoximately and applied to the study
of DIS at small $x$.  We find that the model can naturally
describe the quark distribution in both the large $x$ region and
the small $x$ region dominated by Reggeon exchange.
 \end{abstract}

\phantom{0}
\vspace{7.0in}
\vspace{-6in}
\maketitle

\section{Introduction}

This paper discusses the ingredients of a theory of deep inelastic
scattering (DIS) that holds for {\it all\/} values of the Bjorken
scaling variable $x=Q^2/ 2M_B\nu$ (where $M_B$ is the mass of the
target bound state).    The $Q^2$ of the virtual photon {\it is
fixed at a large value\/}, so the $Q^2$ evolution of the structure
functions will not be discussed.   The individual ingredients are
well known; our emphasis here is on combining these into a unified
picture and in explaining how the various pieces fit together.

At large $Q^2$, the QCD coupling $\alpha_s$ is small, and the
strong interaction dynamics can be treated perturbatively (i.e. to
lowest order)  {\it except in regions where contribution from
higher order loops are enhanced, so that contributions of all
orders must be summed\/}. It is well known that the existence of
the bound state itself (the target $q\bar Q$ system in the simple
example discussed in this paper) depends on the enhancement of
ladder type loops, so the description of the bound state vertex
function must involve nonperturbative physics\footnote{The term
"nonperturbative" is used here in two meanings: firstly, it
implies the summation of infinitely many perturbation diagrams, such as that needed to produce a bound state, and secondly,  as applied to QCD, it implies the involvement of the nonperturbative
confinement mechanism for bound states which cannot be reduced to a finite or infinite perturbation series.}. Nonperturbative
contributions are also known to be important in the region of
small $x$ \cite{Feynman,Ioffe,Yn,Kaidalov}. In the 1960's it was
shown that Regge behavior is generated by infinite sums of ladder
diagrams in the $t$-channel and  explicit  equations  have been
written down for these ladder-type amplitudes \cite{Bertocci,
Simonov} and solved analytically and  numerically  [6].
Furthermore, it was shown in the case of DIS that the finite scaling
limit in $x$ of structure functions with Regge behaviour is
possible to obtain if the Regge residues are properly tuned [7]. In
what follows  we derive explicitly the Regge behaviour of
structure functions in our model and show that it agrees with
general expectations [1-4] and the conjecture made in  Ref.~\cite{Abarb}.

\begin{figure}[t]
  \hspace{2cm}
\includegraphics[width=2.9in]{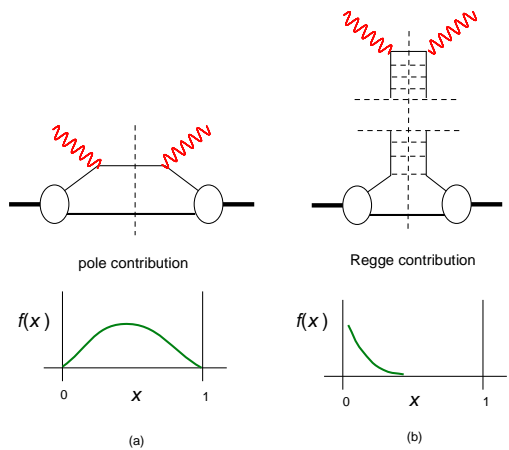}
 \caption{The two contributions to DIS studied in
this paper.  (a) Pole contribution,
important at medium to large $x$.  (b) Regge
contribution arising from the contributions of higher Fock components, which dominates at small $x$.  The
expected qualitative behavior of each contribution is shown below each diagram.\label{fig:DIS}}
\end{figure}

These basic
ideas, and the specific model discussed in this
paper, are illustrated in Fig.~\ref{fig:DIS}.  The two diagrams
show contributions to DIS from a bound state of a heavy anti-quark
(or diquark, if applied to baryons) $\bar Q$ of mass $M$ (assumed
to have no charge, for simplicity) and a light quark $q$ of mass
$m$, interacting through the exchange of scalar gluons ($\phi$
particles with mass $\mu$; confinement is neglected in this
discussion).  To simulate some aspects of the behavior of QCD at
high $Q^2$, the ``quark-gluon'' coupling, $g$, will be assumed to
be small.  Even for small couplings bound states exist (as in
atomic physics, for example); they emerge from the infinite sum of
ladder-like diagrams in the bound state $s$-channel
\cite{grosstext}.  The second diagram, Fig.~\ref{fig:DIS}(b),
describes contributions from the expansion  of higher $q\bar Q
\phi^n$ Fock states, and would normally be suppressed by the
smallness of the coupling constant $g$.  These diagrams are indeed
small, except in the region of large $Q^2$ and small $x$, where
{\it each of the loops in the $t$-channel ladder sum is large, and
contributions of this type must be summed to all orders\/}. This
nonperturbative contribution was discussed in Ref.\
\cite{Simonov}, and shown to exhibt Regge-like behavior.

In accordance with Regge theory [1-4,7], at small $x$ the quark
distributions $F_1$ and $F_2$ should scale as
\bea
W_1(Q^2,x)\to F_1(x)&\simeq &\lambda_1 \left(\frac{1}{x}\right)^{\ell(0)} \label{eq:0}\\
\nu W_2(Q^2,x)\to F_2(x)&\simeq &\lambda_2 \left(\frac{1}{x}\right)^{\ell(0)-1}\label{eq:1}
\eea
where $\ell(0)$ is the intercept of the corresponding Regge
trajectory. The valence quark distributions $f(x)=F_2(x)/x$ are
known to behave at small $x$ as $x^{-0.5}$ while for sea quarks
this behavior is $x^{-1}$, corresponding to the Regge trajectories
of the $\rho$ meson and pomeron respectively  \cite{Kaidalov}.
However the direct matching of the Regge pole contribution to the
scaling behavior at larger $x$ requires a specific tuning of the
coefficients $\lambda_i$, which is not known on general grounds
[2]. The relevant behaviour of these coefficients was suggested in
Ref.~\cite{Abarb} based on the available solutions of scalar
ladder equations with a special relation between the masses
\cite{Nakan} or upper bounds for the massive case \cite{Trei}.  It
would be instructive to use one solvable model with Regge
behavior to understand how the small $x$ behavior explicitly
matches with the behavior at $x$ around one.

This is the primary purpose of the present paper: to show
how these two processes can be  added together to produce a description of DIS valid for {\it all\/} $x$.  We will carry out this discussion using the covariant spectator equation in the ladder
approximation \cite{gross}.

Our treatment of the diagram \ref{fig:DIS}(b) is based on
the integral equation for the imaginary part of the amplitude for
the scattering of two scalar particles of mass $m$ in the  scalar
$\chi^2\phi$ theory, derived independently by similar, but slightly
different methods, in Refs.~\cite{Bertocci} and \cite{Simonov}. From this
equation, which essentially sums up ladder-type diagrams, the
intercept of a Regge trajectory was found  analytically. It
appears that this integral equation is a very useful tool for
studying structure functions, as will be seen in what follows.

This paper is divided into five sections.  In the next section we
derive the integral equation for forward $\gamma q$ elastic
scattering, allowing the external photons to have large $Q^2$,
which is relevant to applications to DIS, and the quark to have a
virtual mass $p^2\ne m^2$, needed for the application to bound
quarks.  In Sec.~III we solve the equation, and in Sec.~IV we add
the photon spin (which was neglected in Sec.~III for simplicity),
and explicily evaluate the two diagrams shown in
Fig.~\ref{fig:DIS}. Our conclusions are presented in Sec.~V and
the Appendices include some further details.

This work can be easily extended to the description of non-forward scattering,
and
can be used to model Generalized Parton Distributions (GPD's)
\cite{Radyushkin,Ji,Diehl:2003ny} which is one of the motivations of the present work,
but these applications are saved for later papers.

\section{Integral equation for photon-quark scattering}

\begin{figure}[b]
\leftline{
\includegraphics[width=3.2in]{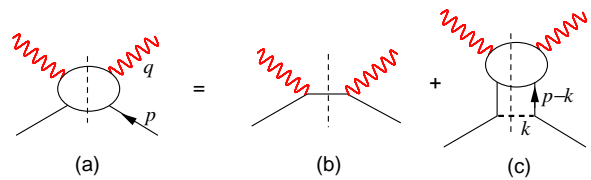}
}
\caption{Diagrammatic representation of
Eq.~(\ref{BSeq2}).  The solid lines are ``quarks'' of mass $m$,
the horizontal dashed line is a ``gluon'' of mass $\mu\to0$, and
the vertical dashed line reminds us that all intermediate
particles are on shell. Fig.(b) is the pole term.\label{fig:qgeq}}
\end{figure}

The integral equation to be derived is illustrated in
Fig.~\ref{fig:qgeq}.  We have specialized the kinematics to
forward scattering, appropriate to applications to DIS, and assume
here that the photon has zero spin (this will be corrected in
Sec.\ IV).  (A more general equation applicable at momentum
transfers $t\ne0$ can be derived using the methods of
Refs.~\cite{Bertocci,Simonov}, and will be discussed elsewhere.)

\begin{figure}[b]
\leftline{\hspace*{-0.7in}
\includegraphics[width=3.2in]{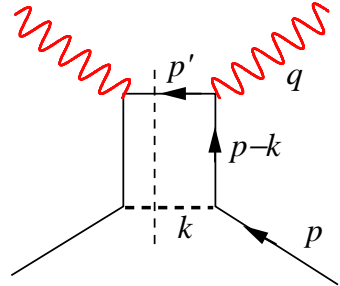}
}
\caption{Diagrammatic representation of box diagram,
Eq.~(\ref{box1}).  The lines are as described in
Fig.~\ref{fig:qgeq}.\label{fig:box}}
\end{figure}

The integral equation can be easily obtained from study of
the box diagram, Fig.~\ref{fig:box}.  The imaginary part of this
amplitude is the total forward cross section for the process
$\gamma+q\to q+g$.  Hence
\bea
{\rm Im}T_2
&=&e^2g^2\int\frac{d^3kd^3p}{(2\pi)^64\omega_kE_{p'}}
\frac{(2\pi)^4\delta^4(k+p'-q-p)}{[m^2-(p-k)^2]^2}\nonumber\\
&=&\frac{e^2g^2}{16\pi^2}\int\frac{d^3k\,
\delta(E_{p'}+\omega_k-p_0-q_0)}{\omega_kE_{p'}\,
[m^2-(p-k)^2]^2}\, , \label{box1}
\eea
where $e$ is the charge of the quark, and
$\omega_k=\sqrt{\mu^2+k^2}$ and $E_{p'}=\sqrt{m^2+p'^2}$.  Since
$q^2=-Q^2<0$, it is convenient to evaluate this integral in the
Breit frame, where
\bea
q&=&(0,0,0,Q)\nonumber\\
p&=&(p_0,0,0,p_z)\, .
\eea
[The external quark is off-shell, and its energy is fixed by
other variables.] Then the integrand is independent of the
azymuthal angle, and the delta function may be used to evaluate
the integral over
$k_\perp^2$, giving
\bea
{\rm Im}T_2(s,x,Q^2)
&=&\frac{e^2g^2}{8\pi}\int_{k^-_{z}}^{k^+_{z}}
\frac{dk_z}{p_0}\frac{1}{[m^2-(p-k)^2]^2}\, ,\quad
\eea
where we have introduced the variables
\bea
s&=&(p+q)^2=p^2+Q^2\left(\frac{1}{x}-1\right)\ge
(m+\mu)^2
\nonumber\\
x&=&\frac{Q^2}{2p\cdot q}=-\frac{Q}{2p_z}\, .
\eea
Note that, because the external quark is off-shell, $x$ is not
restricted, and
\bea
p_0^2=s+Q^2\left(1-\frac{1}{2x}\right)^2\ge0\, .
\eea
The limits on the $k_z$ integral will be discussed below.

It is convenient to write the internal integration variables
in terms of $s'$ and $x'$, where
\bea
s'&=&(p-k+q)^2=m^2=(p-k)^2+Q^2\left(\frac{1}{x'} - 1\right)
\nonumber\\
x'&=&\frac{Q^2}{2(p-k)\cdot
q}=\frac{Q}{2k_z+\frac{Q}{x}}\, .
\eea
This substitution gives the result
\bea
&&{\rm Im}T_2(s,x,Q^2)\nonumber\\
&&\quad=\frac{e^2g^2\,Q}{16\pi}\int_{x_{-}}^{x_{+}}
\frac{dx'}{x'^2}\frac{(s+\eta^2)^{-1/2}}{
[m^2+Q^2\left(\frac{1}{x'}-1\right)-s']^2}\, ,\qquad
\label{box}  \eea
with
\bea
\eta=Q\left(1-\frac{1}{2x}\right)\, .
\eea

The limits on the $x'$ (or $k_z$) integration follow from the
conservation of energy and the requirement that $k_\perp^2\ge0$.
The energy relation is
\bea
\sqrt{\mu^2+k_\perp^2+k_z^2}+\sqrt{s'+k_\perp^2+(\eta-k_z)^2}=p_0
\label{energy}\, .
\eea
This is a quadradic equation in $k_z$, with roots at
\bea
k^\pm_{z}(k_\perp)=\eta
A\pm\sqrt{\left(\frac{\eta^2}{s}+1\right)\left(\frac{\Delta}{4s}-
k_\perp^2\right)}\, ,
\label{limit1}
\eea
where
\bea
A&=&\left(\frac{s-s'+\mu^2}{2s}\right)\nonumber\\
\Delta&=&[s-(\sqrt{s'}+\mu)^2][s-(\sqrt{s'}-\mu)^2]
\, .
\eea
The restrictions on $s$ and $s'$ insure that both $A$ and
$\Delta$ are real and positive.  Introducing $k_z^\pm\equiv
k_z^\pm(0)$, we see that as long as
\bea
k^-_{z}<k_z<k^+_z
\eea
there always exists a value of $k_\perp^2>0$ that satisfies
(\ref{energy}) and the integral is nonzero.  This determines the
limits.  The corresponding limits on $x'$ are
\bea
x_\pm&=&\frac{x}{1+(2xk_z^\mp)/Q}\nonumber\\
&=&\frac{x}{1+(2x-1)A
\mp\frac{x}{sQ}\sqrt{(s+\eta^2)\Delta}}\, .
\eea

It is now a starightforward matter to turn the result (\ref{box})
for the box diagram into an integral equation for the ladder sum.
First observe that the single quark pole term is
\bea
{\rm Im}T_1(s,x,Q^2)
&=&e^2\int
\frac{d^3p'}{(2\pi)^32E_{p'}}(2\pi)^4\delta^4(p'-p-q)
\nonumber\\
&=&2\pi e^2\,\delta(s-m^2)
\, ,\qquad
\label{pole}
\eea
With this notation the box diagram can be rewritten
\begin{widetext}
\bea
{\rm Im}T_2(s,x,Q^2)
=\frac{g^2\,Q}{32\pi^2}\int_{m^2}^{(\sqrt{s}-\mu)^2}
\!\!\!\!\!\!
\frac{ds'}{\sqrt{s+\eta^2}}\int_{x_{-}}^{x_{+}}
\frac{dx'}{x'^2}\frac{{\rm Im}T_1(s',x',Q^2)}{
[m^2+Q^2\left(\frac{1}{x'}-1\right)-s']^2}\, ,\qquad
\label{box2}
\eea
where the upper limit on the $s'$ integral follows from the
general requirement that $\sqrt{s}\ge\sqrt{s'}+\mu$.  The integral
equation we seek is the immediate generalization of (\ref{box2}):
\bea
{\rm Im}T_\infty(s,x,Q^2)
= {\rm Im}T_1(s,x,Q^2) +
\frac{g^2\,Q}{32\pi^2}\int_{m^2}^{(\sqrt{s}-\mu)^2}
\!\!\!\!\!\!
\frac{ds'}{\sqrt{s+\eta^2}}\int_{y_{-}}^{y_{+}}
\frac{dx'}{x'^2}\frac{{\rm Im}T_\infty(s',x',Q^2)}{
[m^2+Q^2\left(\frac{1}{x'}-1\right)-s']^2}\, ,\qquad
\label{BSeq1}
\eea

It is instructive to cast this equation into an alternative form
convenient for solution.  To this end, first introduce virtual mass
parameters for the external and internal quarks:
\bea
\kappa&=&-p^2=Q^2\left(\frac{1}{x}-1\right)-s\nonumber\\
\kappa'&=&-(p-k)^2=Q^2\left(\frac{1}{x'}-1\right)-s'\, .
\eea
For simplicity, we also set $\mu=0$. It is then a
straightforward matter to transform Eq.~(\ref{BSeq1}) into
\bea
{\rm Im}T_\infty(s,\kappa,Q^2)
= {\rm Im}T_1(s,\kappa,Q^2) +
\frac{g^2}{16\pi^2}\int_{m^2}^{s}
\!\!\!\!\!\!
ds'\int_{\kappa'_{-}}^{\kappa'_{+}}
\frac{d\kappa'}{\sqrt{\Delta_0}}
\frac{{\rm Im}T_\infty(s',\kappa',Q^2)}{ [m^2+\kappa']^2} \,
,\qquad
\label{BSeq}
\eea
with
\bea
\kappa'_\pm&=&\kappa+\frac{s-s'}{2s}
\left[Q^2+s-\kappa\pm\sqrt{\Delta_0}\right]\nonumber\\
\Delta_0&=& s^2+\kappa^2+Q^4-2\kappa Q^2+2s\kappa+2sQ^2\, ,
\label{limit1}
\eea
Finally, introduce
new variables
\bea
x_B=\frac{Q^2}{s+Q^2}\qquad\alpha=\frac{s'+Q^2}{s+Q^2}\, .
\eea
In terms of these, Eq.~(\ref{BSeq}) becomes finally
\bea
M(x_B,\kappa,Q^2)
- M_{\rm pole}(x_B,\kappa,Q^2) =
\frac{g^2\,Q^2}{16\pi^2\,x_B}\int_{\alpha_{\rm min}}^{1}
\!\!\!\!\!\!
d\alpha\int_{\kappa_{-}}^{\kappa_{+}}
\frac{d\kappa'}{\sqrt{\Delta_0}}
\frac{M(x_B/\alpha,\kappa',Q^2)}{
[m^2+\kappa']^2}\, ,\qquad
\label{BSeq2}
\eea
\end{widetext}
with
\bea
\alpha_{\rm min}&=&x_B\left(1+\frac{m^2}{Q^2}\right)\nonumber\\
\kappa_\pm&=&\kappa+\frac{1-\alpha}{2(1-x_B)}
\left[\frac{Q^2}{x_B}-\kappa\pm\sqrt{\Delta_0}\right]\nonumber\\
\Delta_0&=& \left(\kappa+\frac{Q^2}{x_B}\right)^2-4\kappa
Q^2\, ,
\label{limit2}
\eea
with Im$T_\infty(s,\kappa,Q^2)=M(x_B,\kappa,Q^2)$, and
\bea
M_{\rm pole}(x_B,\kappa,Q^2)=\frac{2\pi e^2 x_B^2}{Q^2}\;
\delta\left(x_B-\frac{Q^2}{m^2+Q^2}
\right)\, . \label{pole2}
\eea
Equation (\ref{BSeq1}) is exact, and Eq.~(\ref{BSeq2}) is exact
in the $\mu=0$ limit.

\section{The regge limit}

We now apply Eq.~(\ref{BSeq2}) to the case of DIS from an off
shell quark (neglecting the photon spin until the next section).
If the strong coupling $g$ is small, under what circumstances (if
any) will the r.h.s. be larger than the pole term, requiring a
nonperturbative solution to the equation?

\subsection{Approximate equation}

The form of the r.h.s. of (\ref{BSeq2}) suggests that
it will be large only if $x_B$ is small, and hence this term may be
approximated by taking $x_B$ small.  This will be
assumed in what follows, and confirmed later, after the approximate
solution has been obtained.  As $x_B\to0$, the following
approximations hold
\bea
\sqrt{\Delta_0}&\simeq&\frac{Q^2}{x_B}+\kappa+{\cal O}(x_B)
\nonumber\\
\kappa_+&\simeq&\kappa_{\rm max}
+{\cal O}(x_B)
\nonumber\\
\kappa_-&\simeq&\alpha\,\kappa+{\cal O}(x_B)\nonumber\\
\alpha_{\rm min}&\simeq&x_B +{\cal
O}\left(\frac{m^2}{Q^2}\right)\nonumber\\
\kappa_{\rm max}
&\equiv&Q^2\left(\frac{1}{x_B}+1\right)(1-\alpha)+\kappa\, .
\eea
This gives the following
equation, which should hold approximately for all $x_B$ and
$Q^2>\!\!>m^2$
\bea
&&M(x_B,\kappa,Q^2)-M_{\rm pole}(x_B,\kappa,Q^2)\nonumber\\
&&\qquad\quad=
\frac{g^2}{16\pi^2}\int_{x_B}^{1}
\!\!\!\!\!\!
d\alpha\int_{\alpha\kappa}^{\kappa_{\rm max}}\!\!\!
d\kappa'\;\frac{M(x_B/\alpha,\kappa',Q^2)}{[m^2+\kappa']^2}
\, .\qquad \label{BSeq3}
\eea
We emphasize that this equation differs from the exact
Eq.~(\ref{BSeq2}) only in the form of r.h.s., which by
assumption is large only at small $x_B$ where the approximate
form used in (\ref{BSeq3}) is accurate.

\subsection{Approximate solution for small $x_B$}

Following Ref.~\cite{Simonov}, assume that $M$ exhibits a
Regge-like behavior at small
$x_B$
\bea
M(x_B,\kappa,Q^2)\big|_{{\rm small}\,x_B}
\simeq\left(\frac{1}{x_B}\right)^\ell
R(x_B,\kappa,Q^2)\, ,
\eea
where $R$ is finite as $x_B\to0$.  The power $\ell$ will be
determined self consistently from the equation.  Substituting this
anzatz into (\ref{BSeq3}), and neglecting the inhomogeneous pole
term which is small at small $x_B$, gives an equation for
$R(x_B,\kappa,Q^2)$
\bea
&&R(x_B,\kappa,Q^2)\nonumber\\
&&\quad
=
\frac{g^2}{16\pi^2}\int_{x_B}^{1}
\!\!\!\!\!\!
d\alpha\,\alpha^\ell\int_{\alpha\kappa}^{\kappa_{\rm max}}\!\!\!
d\kappa'\;\frac{R(x_B/\alpha,\kappa',Q^2)}{[m^2+\kappa']^2}\,
,\qquad
\label{BSeqR}
\eea
where, by assumption, $R(0,\kappa,Q^2)$ is finite.  This quantity
satisfies the equation
\begin{widetext}
\bea
R(0,\kappa,Q^2)&=& \lambda\,m^2\int_{0}^{1}
d\alpha\,\alpha^\ell\int_{\alpha\kappa}^{\infty}\!
d\kappa'\;\frac{R(0,\kappa',Q^2)}{[m^2+\kappa']^2}\, ,
\nonumber\\
&=&\frac{\lambda\,m^2}{(\ell+1)}\Bigg\{
\int_{\kappa}^{\infty}\!
d\kappa'
+ \int_0^\kappa d\kappa'\left(\frac{\kappa'}
{\kappa}\right)^{\ell+1}
\Bigg\}\,
\frac{R(0,\kappa',Q^2)}{[m^2+\kappa']^2}\, ,\qquad
\label{BSeqR0}
\eea
\end{widetext}
where $\lambda=g^2/(16\pi^2m^2)$ is dimensionless.
Note that $R(0,\kappa,Q^2)$ is independent of $Q^2$.  The
boundary conditions on $R$ are $R\to0$ as $\kappa\to \infty$, and
$R(0,0,Q^2)=$ constant.  Introducing the dimensionless variables
$y=\kappa/m^2$ and $y'=\kappa'/m^2$, and differentiating the
equation by $y$ twice, leads to a differential equation for
$r(y)\equiv R(0,\kappa,Q^2)$
\begin{equation}
 y\, r^{\prime\prime} + (\ell+2)r^{\prime} + {\lambda\over
(y+1)^2}\ r=0.
\label{r0-eq}
\end{equation}
The general solution of this equation is a
hypergeometric function \cite{AS}
\bea
r(y)\propto F\left(a,b,c,\frac{y}{1+y}\right)
\eea
with
\bea
a&=&\frac{1}{2}-\frac{1}{2}\sqrt{1+4\lambda}\nonumber\\
b&=&\frac{1}{2}+\frac{1}{2}\sqrt{1+4\lambda}\nonumber\\
c&=&\ell+2\, . \label{andb}
\eea

The value of $\ell$ is fixed by the $y\to\infty$ boundary
condition, which requires that the hypergeometric
function approach zero as $z\equiv y/(1+y)\to1$.  First, the
hypergeometric series will not converge at $z=1$ unless
$c-a-b>0$, and this requires $\ell+1>0$.  Then, using the
properties of the hypergeometric function
\cite{AS}, we have
\bea
&&F(a,b,c,1)=\frac{\Gamma(c)\Gamma(c-a-b)}{\Gamma(c-a)\Gamma(c-b)}
\nonumber\\
&&=\frac{\Gamma(\ell+2)\Gamma(\ell+1)}
{\Gamma(\ell+\frac{3}{2}+\frac{1}{2}\sqrt{1+4\lambda})
\Gamma(\ell+\frac{3}{2}-\frac{1}{2}\sqrt{1+4\lambda})}
\, .\qquad
\eea
This can be zero if and only if the arguments of one of the
$\Gamma$ functions in the denominator is $-n$, where $n$ is zero or
a positive integer.  Since $\ell+1>0$, it must be the argument of
the second $\Gamma$ function that is $-n$, and hence the only
possible values of $\ell$ are
\bea
\ell=\ell_n\equiv\frac{1}{2}\left(\sqrt{1+4\lambda}-3-2n\right)
\quad n=(0,1,2,
\cdots )\, . \quad\label{lcondition}
\eea
The requirement $\ell+1>0$ yields
\bea
\lambda=\frac{(2\ell+3+2n)^2-1}{4} >0\, ,
\eea
so there are solutions even if $\lambda$ is very small, as
anticipated.  The solutions for $n=0$ were previously
obtained in Ref.~\cite{Simonov}.

For large $\lambda$ there will be
more than one value of $\ell$ that satisfies the condition
(\ref{lcondition}).  The number of values depends on the size of
$\lambda$.  Specifically,
\bea
\begin{array}{ll}
\ell=\ell_0\equiv\frac{1}{2}\left(\sqrt{1+4\lambda}-3\right)
\qquad&{\rm if}
\;\phantom{1}2>\lambda>0 \cr&\cr
\ell=\ell_0\;{\rm or}\;\ell_1
&{\rm if}\;\phantom{1}6>\lambda>2\cr&\cr
\ell=\{\ell_0,\ell_1,\ell_2\}&
{\rm if}\;12>\lambda>6\cr\cdots& \label{l0sol}
\end{array}
\eea
For $\lambda>2$ the general small $x_B$ solution will be a sum of
contributions from the different trajectories
\bea
M(x_B,\kappa,Q^2)\big|_{{\rm small}\,x_B}
=\sum_{n=0}^{n_{\rm max}}
c_n\left(\frac{1}{x_B}\right)^{\ell_n}\!\!\!
r_n\left(\frac{\kappa}{m^2}\right)\, .
\eea
The solution $r_0(y)$ for the ``leading'' trajectory, $n=0$, is
\bea
r_0(y)&=&F(a,b,b,z)=\left(\frac{1}{1-z}\right)^a
=\left(\frac{m^2}{m^2+\kappa}\right)^{\ell_0+1} \qquad
\label{r0-solution}
\eea
For small $\lambda$ (i.e. $\lambda<2$) there is only one
trajectory, and we obtain the unique approximation
\bea
M_0(x_B,\kappa,Q^2)
=c_0\left(\frac{1}{x_B}\right)^{\ell_0}\!\!\!
\left(\frac{m^2}{m^2+\kappa}\right)^{\ell_0+1}\!\!\! ,\quad
\label{sol}
\eea
with $c_0$ an undetermined coefficient.

Some additional arguments which demonstrate that we have found all of the solutions are given in Appendix A

At this point we are left with several unanswered questions: (i)
How is the coefficient $c_0$ to be fixed? (ii) Only the
leading dependence at small $x_B$ is given by $M_0$.  How
large are the nonleading terms at small $x_B$ and what is their
analytic form?  (iii) How is the pole term to be included in the
solution and how is the large $x_B$ dependence described
approximately?

All of these questions will be answered in the following subsection.
 It turns out that substitution of the
ansatz
\bea
M(x_B,\kappa,Q^2) =M_{\rm pole}(x_B,\kappa,Q^2)
+ M_0(x_B,\kappa,Q^2)\, ,\qquad
\label{sol2}
\eea
into Eq.~(\ref{BSeq3}) will show how to
fix the coefficient $c_0$ so that (\ref{sol2})
will give a satisfactory solution to (\ref{BSeq3}) for {\it all\/}
$x_B$.

\subsection{Approximate solution for all $x_B$}

Substituting $M_0$ into the r.h.s. of the asymptotic Eq.
(\ref{BSeq3}) gives
\begin{widetext}
\bea
{\rm r.h.s.}&=&
c_0 \lambda\,m^2\int_{x_B}^{1}
\!\!\!\!\!\!
d\alpha\left(\frac{\alpha}{x_B}\right)^{\ell_0}
\int_{\alpha\kappa}^{\kappa_{\rm max}}\!\!\!
d\kappa'\;\frac{m^{2\ell_0+2}}{[m^2+\kappa']^{\ell_0+3}}
\qquad\nonumber\\
&=&\frac{c_0 \lambda}{(\ell_0+2)}\int_{x_B}^{1}
\!\!\!\!
d\alpha\left(\frac{\alpha}{x_B}\right)^{\ell_0}
\Bigg\{\left(\frac{m^2}{m^2+\alpha\kappa}\right)^{\ell_0+2}
-
\left(\frac{m^2}{m^2+\kappa_{\rm max}}\right)^{\ell_0+2}\Bigg\}
\, .
\label{rhs}
\eea
The first term is the leading term, and doing the $\alpha$
integration gives
\bea
{\rm r.h.s.}\big|_{\rm 1st}&=&\frac{c_0
\lambda}{(\ell_0+2)(\ell_0+1)}\Bigg\{\frac{1}{x_B^{\ell_0}}
\left(\frac{m^2}{m^2+\kappa}\right)^{\ell_0+1}
-x_B\left(\frac{m^2}{m^2+x_B\kappa}\right)^{\ell_0+1}\Bigg\}
\quad\nonumber\\
&=& M_0(x_B,\kappa,Q^2) -c_0\;x_B
\left(\frac{m^2}{m^2+x_B\kappa}\right)^{\ell_0+1}\, .
\label{rhsfirst}
\eea
This shows that $M_0$ is a
solution of the {\it homogeneous\/} equation with an error [the
second term in (\ref{rhsfirst})]  which vanishes linearly as
$x_B\to0$, showing that $M_0$ is accurate to
${\cal O}(x_B^{1+\ell_0})$.  This is not very satisfactory,
particularly when $\lambda$ is small and $\ell_0\to-1$.  The pole
term, not yet discussed, will correct this problem. The second term
in (\ref{rhs}) gives a correction proportional to
$m^2x_B/Q^2$.  At large $Q^2$ it becomes
\bea
\lim_{Q^2\to\infty}{\rm r.h.s.}\big|_{\rm 2nd}&\to&
\bigg[M_0(x_B,\kappa,Q^2)
-c_0\left(\frac{m^2x_B^2}{(m^2+\kappa) \,x_B+
Q^2(1-x_B^2)}\right)^{\ell_0+1}
\bigg]\frac{m^2x_B}{Q^2(1+x_B)}\, .
\label{rhssecond}
\eea
In applications to DIS, these terms can safely be neglected.

Next, we look at the effect of inserting the pole term
(\ref{pole2}) into the r.h.s. of (\ref{BSeq3}).  This gives
\bea
{\rm r.h.s.}\big|_{\rm pole}&=&
\frac{2\pi\lambda\,e^2\,x_B^2}{Q^2}\int_{x_B}^{1}
\!\!
\frac{d\alpha}{\alpha^2}\;
\delta\left(\frac{x_B}{\alpha}-\frac{Q^2}{m^2+Q^2}\right)
\int_{\alpha\kappa}^{\kappa_{\rm max}}\!\!\!
d\kappa'\;\frac{m^{2}}{[m^2+\kappa']^{2}}
\qquad\nonumber\\
&\simeq&\frac{2\pi\lambda\,e^2\,x_B}{Q^2}
\Bigg\{\frac{m^2}{m^2+x_B\kappa} -
\frac{m^2x_B}{(m^2+\kappa)\,x_B+Q^2(1-x_B^2)}\Bigg\}
\, ,
\label{rhspole}
\eea
where we assumed that $Q^2>\!\!>m^2$ so that the delta function
gives $\alpha=x_B$.  The second term is small at large $Q^2$
(except at the exceptional point $x_B=1$).  Combining the first
term in (\ref{rhspole}) with the correction term in
(\ref{rhsfirst}) gives
\bea
{\rm r.h.s.}\big|_{{\rm pole}+ {\rm corr}}&=&
\frac{2\pi\lambda\,e^2\,x_B}{Q^2}
\left(\frac{m^2}{m^2+x_B\kappa} \right)
-c_0\;x_B
\left(\frac{m^2}{m^2+x_B\kappa}\right)^{\ell_0+1}\nonumber\\
&\simeq&x_B\;\left[\frac{2\pi\lambda\,e^2}{Q^2}-c_0\right]
-x_B^2\;\left[\frac{2\pi\lambda\,e^2}{Q^2}-(\ell_0+1)\,c_0\right]
\frac{\kappa}{m^2}+\cdots
\, ,
\label{rhscombined}
\eea
\end{widetext}
Setting
\bea
c_0=\frac{2\pi\lambda\,e^2}{Q^2} \label{c0choice}
\eea
will cancel the leading error term and insure that the ansatz
(\ref{sol2}) is accurate to ${\cal O}(x_B^{2+\ell_0})$ and
vanishes as $x_B\to0$ for all $\ell_0$.    We conclude that the
solution (\ref{sol2}) with the choice (\ref{c0choice}) is accurate
for all $x_B$ and $Q^2$ as long as $Q^2>\!\!>m^2$.

 \subsection{Summation of the series for small $x_B$ and $\kappa$}

Before leaving this discussion, it is instructive to show
explicitly how the Regge behavior (\ref{sol}) arises by summing
the infinite number of terms in the series defined by the
iteration of  Eq.~(\ref{BSeq3}).

Iteration of Eq.~(\ref{BSeq3}) generates the following series:
\bea
&&M(x_B,\kappa,Q^2)-M_{\rm pole}(x_B,\kappa,Q^2)\nonumber\\
&&\qquad\qquad\qquad=\sum_{n=0}^\infty M^{(n+1)}(x_B,\kappa,Q^2)
 \label{BSeqseries}
\eea
where, taking $\kappa_{\rm max}\to\infty$,
\bea
 M^{(n+1)}(x_B,\kappa,Q^2)=
 \int_{\alpha,\kappa'} \frac{M^{(n)}(x_B/\alpha,\kappa',Q^2)}{[m^2+\kappa']^2}
\, ,  \label{BSeqseries1}
\eea
with
 \bea
 M^{(0)}(x_B,\kappa,Q^2)\equiv
 M_{\rm pole}(x_B,\kappa,Q^2)
 \eea
 and
 \bea
 \int_{\alpha,\kappa'}\equiv\frac{g^2}{16\pi^2}\int_{x_B}^{1}
\!\!\!\!\!\!
d\alpha\int_{\alpha\kappa}^{\infty}\!\!\!
d\kappa' \, .
 \eea

Near $\kappa=0$, the leading $\log(x_B)$ terms in this
series can be summed.  In Appendix B we show that these leading terms are given by the simpler series
 \bea
 M^{(n+1)}_{\rm leading}(x_B)=\lambda
 \int_{x_B}^1d\alpha\;  M^{(n)}_{\rm leading}(x_B/\alpha)
\, .  \label{BSeqseries2}
 \eea
 With the aid of the identity
 \bea
 \int_1^{1/x_B} \frac{dy}{y}\log^n(y)=\frac{1}{n+1}\log^{n+1} \left(\frac{1}{x_B}\right)\, ,
 \eea
 the general term becomes
 \bea
  M^{(n+1)}_{\rm leading}(x_B)=\frac{c_0x_B}{n!}\left[\lambda\log\left(\frac {1}{x_B}\right)\right]^n\, ,
 \eea
 and summing to all orders gives
 \bea
c_0x_B \sum_{n=0}^\infty
 \frac{1}{n!}\left[\lambda\log\left(\frac {1}{x_B}\right)\right]^n=c_0x_B\left(\frac{1}{x_B}\right)^\lambda
\eea
in precise agreement with the small $\lambda$ and small $\kappa$ limit of (\ref{sol}) (where $\ell_0\simeq -1+\lambda$).

This calculation confirms the normalization the solution, but it
also shows explicitly that the {\it Regge behavior arises when
$\lambda \log x_B\sim -1$, the condition that insures that all of
the loops are of comparable size\/}.  Under these conditions
perturbation theory breaks down and a nonperturbative summation of
all diagrams is needed.

\section{dis structure functions}

In this section the results of the previous section are applied to
DIS.  First we add the photon vector degrees of freedom, then we
compute the two diagrams shown in Fig.~\ref{fig:DIS}.

\subsection{Photon degrees of freedom}

The vector degrees of freedom of the photon will introduce
additional structure into the pole term (\ref{pole2}) and the
scattering amplitude (\ref{sol2}).  The pole term becomes
\bea
&&J_{\rm pole}^{\mu\nu}(x_B,p,Q)\nonumber\\
&&\qquad = (2p-q)^\mu(2p-q)^\nu
M_{\rm pole}(x_B,-p^2,Q^2)\, . \qquad\quad \label{vecpole}
\eea
If both the initial and final quarks were on-shell, this would be
gauge invariant, but the initial quark is not on-shell.  We impose
gauge invariance in the manner suggested in Ref.~\cite{BG1}, where
it was found that the substitution
\bea
J^\mu\to J^\mu-\frac{J\cdot q\;q^\mu}{q^2}
\eea
insured that the final state interactions vanished in the DIS
limit.  This choice therefore preserves the dominance of the pole
term in the DIS limit, and is what is needed in the absence of a
full theory.  Applied to the product of two currents, it leads to
the replacement
\bea
J^{\mu\nu}&\to& J^{\mu\nu}-\frac{(J^{\mu\nu'}
q_{\nu'})\; q^\nu}{q^2} -\frac{(J^{\mu'\nu} q_{\mu'}) \;q^\mu}{q^2}
\nonumber\\
&&+\frac{(J^{\mu'\nu'}q_{\mu'} q_{\nu'})\;q^\mu q^\nu }{q^4}\, .
\eea
This, in turn, give the substitutions
\bea
&&g^{\mu\nu}\to g^{\mu\nu}-\frac{q^\mu\,q^\nu}{q^2} \equiv
\tilde g^{\mu\nu}
\nonumber\\
&&p^\mu p^\nu\to\left(p^\mu-\frac{p\cdot q\,q^\mu}{q^2}\right)
\left(p^\nu-\frac{p\cdot q\,q^\nu}{q^2}\right)
\equiv\tilde p^\mu \tilde p^\nu
\nonumber\\
&&q^\mu q^\nu\to0\nonumber\\
&&p^\mu q^\nu + q^\mu p^\nu\to0\, , \label{iden1}
\eea
and the pole term (\ref{vecpole}) is transformed into the familiar
form
\bea
&&J_{\rm pole}^{\mu\nu}(x_B,p,Q) = 4\tilde p^\mu \tilde p^\nu
M_{\rm pole}(x_B,\kappa,Q^2)\, , \qquad\quad
\eea
where, as before, $\kappa=-p^2$.
The general structure of the full amplitude will be expressed in
term of two scalar functions
\bea
J^{\mu\nu}(x_B,p,Q) = &-&\tilde g^{\mu\nu}\; w_1(x_B,\kappa,Q^2)
\nonumber\\
&+&\frac{\tilde p^\mu\tilde p^\nu}{\tilde p^2}
\; w_2(x_B,\kappa,Q^2)\, , \qquad \label{w1andw2}
\eea
where the $w_i$ differ from the familiar $W_i$ of DIS by
normalization and the dependence on the off-shell quark mass.

When the form (\ref{w1andw2}) is inserted into the integral that
leads eventually to  Eq.~(\ref{BSeq2}) or (\ref{BSeq3}), the
integration over the azymuthal angle $\pi$, which was originally
trivial, now requires use of the identity
\bea
&&\frac{1}{2\pi}\int_0^{2\pi} d\phi\;(\tilde p-\tilde k)^\mu
(\tilde p-\tilde k)^\nu\nonumber\\
&&\qquad\to-\frac{1}{2}k_\perp^2 \tilde g^{\mu\nu}
+\frac{\tilde p^\mu \tilde p^\nu}{\tilde p^2}\left(
\frac{1}{2}k_\perp^2+(p_0-k_0)^2\right)\, .\qquad \label{iden2}
\eea
The quantities in this equation can be expressed in terms of the
variables used in Eq.~(\ref{BSeq3}).  In the small $x_B$, large
$Q^2$ limit
\bea
&&k_\perp^2\to(1-\alpha)\,(\kappa'-\alpha\kappa)\nonumber\\
&&\tilde p^2=p_0^2\to\frac{Q^2}{4x_B^2}\nonumber\\
&&\frac{1}{2}k_\perp^2+(p_0-k_0)^2\to\frac{\alpha^2 Q^2}{4x_B^2}
\nonumber\\
&&(\tilde p-\tilde k)^2\to\frac{\alpha^2 Q^2}{4x_B^2}
\eea
\begin{widetext}
Hence Eq.~(\ref{BSeq3}) generalizes to a system of coupled
equations for the $w_i$
\bea
&&w_1(x_B,\kappa,Q^2)=
\frac{g^2}{16\pi^2}\int_{x_B}^{1}
\!\!\!\!\!\!
d\alpha\int_{\alpha\kappa}^{\kappa_{\rm max}}\!\!\!
\frac{d\kappa'}{[m^2+\kappa']^2}\Big\{\frac{2x_B^2\,
(1-\alpha)\,(\kappa'-\alpha\kappa)}{\alpha^2Q^2}\,
w_2(x_B/\alpha,\kappa',Q^2) + w_1(x_B/\alpha,\kappa',Q^2) \Big\}
\nonumber\\
&&w_2(x_B,\kappa,Q^2)-4\tilde p^2 M_{\rm pole}(x_B,\kappa,Q^2)=
\frac{g^2}{16\pi^2}\int_{x_B}^{1}
\!\!\!\!
d\alpha\int_{\alpha\kappa}^{\kappa_{\rm max}}\!\!\!
d\kappa' \;\frac{
w_2(x_B/\alpha,\kappa',Q^2)}{[m^2+\kappa']^2}
\, .\qquad
\label{eqwi}
\eea
\end{widetext}
Note that the equation for $w_2$ is identical to (\ref{BSeq3})
except for the factor of $4\tilde p^2$ multiplying the pole term,
and that $w_1$ is smaller than $w_2$ by $Q^{-2}$, and can be
ignored.  The factor of $4\tilde p^2$ changes the value of $c_0$
extracted from the cancellation described in Eq.\
(\ref{rhscombined}), but otherwise does not alter the argument.
Incorporating these modifications into the solution obtained in
the previous section, and taking the high $Q^2$ limit, gives
\bea
w_1(x_B,\kappa,Q^2)&\to& 0
\nonumber\\
w_2(x_B,\kappa,Q^2)&=&2\pi e^2\, \delta(x_B-1)
\nonumber\\
&&+ \frac{ e^2g^2}{8\pi m^2}
\left(\frac{1}{x_B}\right)^{\ell_0}\!\!\!
\left(\frac{m^2}{m^2+\kappa}\right)^{\ell_0+1}\!\!\!\!\!\!.\qquad
\label{solw2}
\eea
Note that this solution is dimensionless and independent of $Q^2$.

\subsection{Structure function for scalar bound states}

\begin{figure}
\leftline{
\includegraphics[width=2.5in]{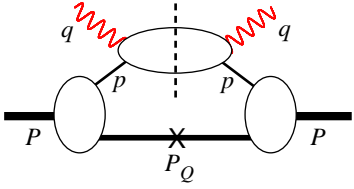}
}
\caption{Diagram showing the convolution of the scattering
amplitude $w_2$ from Eq.~(\ref{solw2}) with the covariant bound
state vertex function, as canculated in the spectator theory.  The
on-shell heavy quark is labeled by an $\times$ in the diagram.
\label{fig:DIS2}}
\end{figure}

We now use the result (\ref{solw2}) to construct the contributions
from each of the diagrams shown in Fig.~\ref{fig:DIS}.  Figure
\ref{fig:DIS2} shows the labeling of momenta in the generic
convolution diagram.  In this part of the
calculation we choose a different coordinate system, with momenta
\bea
P&=&\{M_B,{\bf 0},0\}
\nonumber\\
P_Q&=&\{{\cal E},-{\bf p}_\perp,-p_3\}
\nonumber\\
q&=&\{\nu,{\bf 0},q_{_L}\}\nonumber\\
p&=&\{M_B-{\cal E},{\bf p}_\perp,p_3\}\, ,
\label{mom2}
\eea
where ${\cal E}=\sqrt{M^2+p^2_\perp+p_3^2}$ as required by the
condition that the heavy quark is on its positive energy
mass-shell.  Since the off-shell quark structure function
$w_2$ is manifestly covariant, it may be written in any coordinate
system, and we will use this property to write it in the coordinate
system defined by the choices given in Eq.~(\ref{mom2}). With this
choice the variable $x$ takes on the familiar form
\bea
x=\frac{Q^2}{2M_B\nu}\, .
\eea

In the covariant spectator formalism\cite{gross}, the diagram in
Fig.~\ref{fig:DIS2} is
\bea
J_B^{\mu\nu}(P,q)&=&\int\frac{d\phi\, dp_\perp^2 dp_3}{(2\pi)^3
4{\cal E}}\;\Psi^2(p^2)\,J^{\mu\nu}(x_B,p,q) \nonumber\\
&\equiv& \int d\Sigma \; J^{\mu\nu}(x_B,p,q)
\label{JB}
\eea
where $J^{\mu\nu}$ was defined in Eq.~(\ref{w1andw2}) and
\bea
\Psi(p^2)=\frac{\Gamma(p^2)}{m^2-p^2} \label{relwf}
\eea
is the relativistic wave function (by definition), with
$\Gamma(p^2)$ the covariant vertex function for the coupling of a
bound meson with mass $M_B$ to an on-shell heavy quark with mass
$M$ and an off-shell light quark with four-momentum $p$.  In the
scalar theory we are describing this can be a function of $p^2$
only.

Decomposing (\ref{JB}) into the standard scalar functions
\bea
J^{\mu\nu}_B(P,q)=-W_1(\nu,Q^2)\,\tilde
g^{\mu\nu}+W_2(\nu,Q^2)\frac{\tilde P^\mu \tilde P^\nu}{M_B^2}\, ,
\eea
multiplying by $g_{\mu\nu}$ and $P_\mu P_\nu$, and separating $W_1$ and $W_2$, gives the following expressions
\bea
\left(1+\frac{\nu}{2M_B x}\right)W_2(\nu,Q^2)=\frac{1}{2}\int d \Sigma\; w_2
\left[3\,\theta -1\right]
\nonumber\\
W_1(\nu,Q^2)=\int d\Sigma \;\bigg\{ w_1 + \frac{1}{2}w_2\left[\theta -1\right]\bigg\}\, ,   \qquad
\label{Bigw1&w2}
\eea
with
\bea
w_i&=& w_i(x_B,-p^2,Q^2)
\nonumber\\
\theta&=&\theta(p_\perp^2,p_3)=\frac{(P\cdot\tilde p)^2}{\tilde p^2 M_B^2 \left(1+\frac{\nu}{2M_B x}\right)}\, ,
\eea
and we retained $w_1$ for completeness. In the high $Q^2$ limit,
$q_{_L}\sim\nu\sim Q^2$, and $P\cdot \tilde p\to (P\cdot q)(p\cdot
q)/Q^2$ and $\tilde p^2\to (p\cdot q)^2/Q^2$ giving
\bea
\lim_{Q^2\to\infty} \theta(p_\perp^2,p_3)\to 1\, .
\eea
Hence $W_1$ vanishes (as expected for scalar constituents), and, as $Q^2\to\infty$,
\bea
\nu W_2(\nu,Q^2)\to2M_B x\int\frac{\, dp_\perp^2 dp_3}{(2\pi)^2
4{\cal E}}\;\Psi^2(p^2)\, w_2 \label{finalW2}
\eea

The integral (\ref{finalW2}) is best expressed in terms of the
independent variables $p_\perp^2$ and $p_-$.  Introducing the
dimensionless variable $z=p_-/M_B$, so that
\bea
-p^2&=&\frac{1}{1-z}\left(p_\perp^2+zM^2-z(1-z)M_B^2\right)
\nonumber\\
x_B&=&\frac{Q^2}{s+Q^2}\to \frac{x}{z} \;\;\mbox{(at large $Q^2$)}
\, ,
\eea
and introducing the constant
\bea
S_0=m^2M_B\frac{e^2}{4\pi}
\eea
gives
\begin{widetext}
\bea
\nu W_2(\nu,Q^2)&=&x\,S_0 \int_0^\infty \frac{d p_\perp^2}{m^2}
\int_x^1\frac{dz}{1-z}\; \Psi^2(p^2)\left[
\delta\left(\frac{x}{z}-1\right) + \lambda\left(\frac{z}{x}
\right)^{\ell_0}
\left(\frac{m^2}{m^2-p^2}\right)^{\ell_0+1}\right]
\, . \label{solW2}
\eea
\end{widetext}
Note that $\nu W_2$ goes like $x^{1-\ell_0}$ at small $x$, as it
is expected on general grounds [1-4].

\begin{figure}[b]
\leftline{
\includegraphics[width=3in]{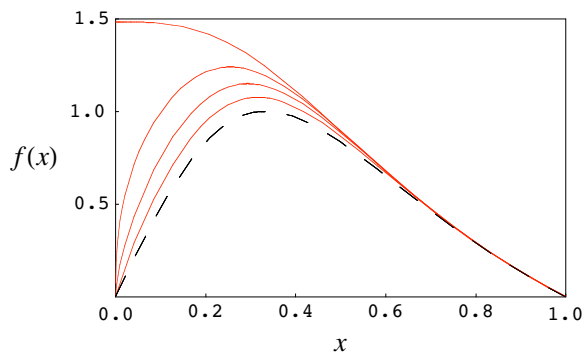}
}
\caption{The structure function $f(x)=\nu W_2/x$ as a function of $x$
for the model discussed in the text.  The solid lines are the sum
of the pole and the Regge contributions for values of
$\lambda=0.2, 0.5, 1.0$ and
$2.0$, with the curves increasing in value as $\lambda$
increases.  The dashed line is the pole
contribution, which is independent of $\lambda$ in our simple
approximation.
\label{fig:results}}
\end{figure}

It is instructive and amusing to evaluate $\nu W_2$ for a simple
model, and look at the comparative behavor of the two terms.  A
detailed study would require the solution of the  bound state
spectator equation for the spectator wave function.  While this is
not difficult, it is also not necessary in order to obtain a
qualitative understanding of the physics. To this end we introduce
a very simple ansatz for the wave function
\bea
\Psi(p^2)&=&\left(\frac{m^2}{m^2-p^2}\right)^\beta\nonumber\\
&=&(1-z)^\beta\left(\frac{m^2}{{\cal
M}^2(z)+p_\perp^2}\right)^\beta \label{model}
\eea
with
\bea
{\cal M}^2(z)=(1-z)\,m^2+z\,M^2-z(1-z)M_B^2
\eea
In the following example we take $\beta=1$ for simplicity.

We now evaluate each of the terms in (\ref{solW2}).  The
first term is the pole contribution, shown in
Fig.~\ref{fig:DIS}(a).  Immediate evaluation of the $z$ integral
gives
\bea
\nu W_2\Big|_{\rm pole}=S_0\,\frac{x^2}{1-x}\int_0^\infty
\frac{dp_\perp^2}{m^2}\;\Psi^2(p^2(z=x))\, ,
\eea
and evaluating the $p_\perp$ integral gives
\bea
\nu W_2\Big|_{\rm
pole}=\frac{S_0\,x^2\,(1-x)}{1-x+x\,M_0^2-x(1-x)M_{B0}^2}\, .
\eea
This function is shown (as a dashed line) in
Fig.~\ref{fig:results}, for the choices $M_0\equiv M/m=2$,
$M_{B0}\equiv M_B/m=2$, and $S_0=10$.  It gives the familar shape anticipated in the drawing shown in
Fig.~\ref{fig:DIS}; it is zero at the $x=0$ and $x=1$ boundaries.

The new result of this paper is the inclusion of the small $x$
Regge term with  a normalization consistent with the valance
contribution.  For the model (\ref{model}) this term becomes
\bea
\nu W_2\Big|_{\rm Regge}=\lambda
\left(\frac{1}{x}\right)^{\ell_0-1} G(x,\lambda) \label{Reg1}
\eea
with
\bea
G(x,\lambda)=\frac{S_0}{\ell_0+2}
\int_x^1dz\,z^{\ell_0}
\left(\frac{m^2(1-z)}{{\cal M}^2(z)}\right)^{\ell_0+2}
\eea
where the power $\ell_0$ was given in Eq.~(\ref{l0sol}).  [Note
that Eq.\ (\ref{Reg1}) agrees with the general result of Eq.\
(\ref{eq:1}).]   Since $1-\ell_0\le2$, the Regge term is always
larger than the pole term near $x=0$, but the importance of
this term depends on the size of $\lambda$.   If $\ell_0>0$
($\lambda> 2$) the Regge contribution to the quark distribution
$f(x)=\nu W_2/x$ is singular as $x\to0$, and the structure
function has the shape shown in Fig.\ \ref{fig:DIS}.  But, even
for small values of $\lambda$, Fig.\ \ref{fig:results} shows that
the Regge term enhances the distribution significantly.

\section{conclusions}


This paper applies the Regge concepts originally developed in
Ref.~\cite{Simonov} to DIS.  We find that an approximate solution
can be obtained that can explain, qualitatively at least, the
enhancement of the structure function at small $x$.  The physics
that produces this effect is the inelastic production of any
number of massless gluons when $x$ is small, and $s$ is
corresponding large.

We emphasize that this calculation shows that, even when the
strong coupling is small, {\it so that perturbation theory should
apply over most of the phase space\/}, there are regions when an
an infinite number of diagrams must be summed to get the correct
result.  The need to do this for the description of bound states
is familar to most physicists.  Here we have discussed  it is also
necessary in the small $x$ region when the production of massless
gluons must be summed to all orders to give the correct result.

The scalar theory examined in this paper gives an unrealistically
small result $\ell(0)\simeq-1+\lambda$ for the $t=0$ intercept of
the Regge trajectory.  A realistic treatment based on QCD,
including spin and confinement, should give an intercept of
$\ell(0)=1/2$, corresponding to the $\rho$ trajectory (the largest
trajectory possible for a $q\bar q$ ladder).  While this result
could be mimicked in this model by choosing $\lambda= 3.75$, this
is too large to be realistic, and the failure is due to the
physics missing from the scalar theory.

The gluon distribution, which corresponds to $\ell(0)\simeq1$, is expected to arise from single and multiple pomeron exchanges \cite{Kaidalov}, mechanisms not included in the model discussed here.  The calculation of  ladder exchanges of massless
gluons of spin one is known to give an intercept around one for small coupling constant, in agreement with the small $x$ behavior of structure functions.

There are many interesting issues and unsolved  problems yet to be dealt with.   The very notion of Regge trajectories $\alpha_i(t)$ in QCD is understood to some extent only for positive $t$ \cite{Kaidalov}, where it has been shown that the
 trajectories are indeed linear and the intercepts $\alpha_i(0)=
\ell_i(0)$  have been explicitly calculated theoretically  \cite{Badalian}.  At the same time, the behavior of $\alpha_i(t)$ for $t<0$ in QCD (and other field theory models) is still unknown, though very interesting from both experimental and
 theoretical point of view.

As trajectories for $t<0$ are understood, it will be of interest to apply these ideas to other processes, such as exclusive DIS and Virtual Compton Scattering, where $t \neq 0$ and $Q^2 \neq {Q^\prime}^2$.  Such applications will yield models for the interesting Generalized Parton Distributions (GPDs).  Another point of interest is the interconnection of Regge behavior and the DGLAP evolution of structure functions.  All of this will require an explicit understanding of the behavior of the Regge trajectories at both positive and negative values of $t$.  To obtain linear Regge trajectories with realistic intercepts, and to describe structure functions quanitatively in the whole $x$ region, one needs to modify the ladder kernel of the equations used in this paper to  include the nonperturbative effects arising from confinement.
Study of these fundamental and difficult problems is planned for future work.

\acknowledgements

This work was supported in part by the US Department of Energy
under grant No.~DE-FG02-97ER41032. The Southeastern Universities
Research Association (SURA) operates the Thomas Jefferson National
Accelerator Facility under DOE contract DE-AC05-84ER40150.
One of the authors (Yu.S.) is grateful for useful discussions to
 K.G.Boreskov, A.B.Kaidalov and O.V.Kancheli.

\appendix

\section{}


In this appendix we give an alternative demonstration that study
of Eq.~(\ref{BSeqR}) near $x=0$ will not produce solutions that
have not already been considered.

Rewrite Eq.~(\ref{BSeqR})
\begin{equation}
 r(y,x) = \lambda
              \int\limits_x^1 d\alpha \ \alpha^l
              \int\limits_{y\alpha}^\infty dy^\prime
               { r(y^\prime, {x / \alpha}) \over
                   (y^\prime +1)^2 }.
\label{r-eq}
\end{equation}
where use the fact that the upper limit of the $y'$ integration
can be taken to infinity with errors that vanish as $Q^2\to
\infty$.

 To simplify the form of the integral in this equation,
 let us define a new function $\phi(y,x)$ trough
\begin{equation}
 r(y,x) = \phi(y,x)(y+1)^2 x^{l+1}.
\end{equation}

 The equation for $\phi$
\begin{equation}
 \phi(y,x) (y+1)^2 = \lambda
              \int\limits_x^1 { d\alpha \over \alpha}
              \int\limits_{y\alpha}^\infty dy^\prime
               \phi \left (y^\prime, {x \over \alpha} \right )
\label{phi-eq}
\end{equation}
 may be converted to a partial derivative equation through the
 following steps.

 First, after taking $y$- derivative from both sides
 in Eq.(\ref{phi-eq}), we obtain
\begin{equation}
 {\partial \left ( \phi(y,x) (y+1)^2 \right )
  \over \partial y }
                    = - \lambda
              \int\limits_x^1 d\alpha
               \phi \left (y\alpha, {x \over \alpha} \right ).
\label{dphi-eq}
\end{equation}
 Changing the integration variable from $\alpha$ to
 $z={\alpha\over x}$, we write Eq.(\ref{dphi-eq}) as
\begin{equation}
 \Phi_y={\partial \over \partial y}
       \left ( \phi(y,x) (y+1)^2 \right )
                    = - \lambda x
              \int\limits_1^{1/x} dz
               \phi \left (xyz, {1 \over z} \right ),
\label{Phi-y}
\end{equation}
 where the notations
 $$\Phi=\phi(y,x) (y+1)^2 = r(y,x) x^{-(l+1)}, \ \ \
 \Phi_y\equiv{\partial \over \partial y} \Phi$$
 are used.

 For partial $x$ and $y$ -- derivatives of $\Phi_y$ Eq.(\ref{Phi-y}),
 we have
\begin{widetext}
\begin{eqnarray}
 \Phi_{y,x}&\equiv&{\partial \over \partial x} \Phi_y
      = {1 \over x} \Phi_y
       - \left ( -{1 \over x^2} \right ) \lambda x \phi(y,x)
       - \lambda x \int\limits_1^{1/x} dz \ yz
           \left .
            \left ( {\partial\phi\left (\xi, {1 \over z}\right )
                          \over \partial \xi } \right )
           \right \vert_{\xi=xyz},
\label{Phi-yx} \\
 \Phi_{y,y}&\equiv&{\partial \over \partial y} \Phi_y
       = - \lambda x
              \int\limits_1^{1/x} dz \ xz
           \left .
            \left ( {\partial\phi\left (\xi, {1 \over z}\right )
                          \over \partial \xi } \right )
           \right \vert_{\xi=xyz}.
\label{Phi-yy}
\end{eqnarray}

 The integral in the last term in Eq.(\ref{Phi-yx}) is the same as
in
 Eq.(\ref{Phi-yy}), therefore Eq.(\ref{Phi-yx}) may be re-written as
\begin{equation}
 \Phi_{y,x}= {1\over x} \Phi_y +  {\lambda\over x} \phi(y,x) +
            {y\over x} \Phi_{y,y},
\label{Phi-eq1}
\end{equation}
 or, finally,
\begin{equation}
 y \Phi_{y,y} + \Phi_y +  {\lambda\over (y+1)^2 } \Phi - x\Phi_{y,x}=0.
\label{Phi-eq}
\end{equation}
\end{widetext}


 The form of the $\Phi_{y,x}$ -- term in Eq.(\ref{Phi-eq}) suggests
 a direct way to separate the variables. A simple observation is
 that for a function of the form $F(y,x)=f(y) x^n$ with arbitrary
$n$,
 the last term in Eq.(\ref{Phi-eq}) with mixed derivatives becomes
 a term with only $y$ -- derivative:
\begin{equation}
 - xF_{y,x} = -nF_{y}.
\end{equation}
 Therefore, we will look for a solution of Eq.(\ref{Phi-eq})
 in the form
\begin{equation}
 \Phi(y,x) = \sum_n c_n x^n r_n(y) x^{-(l+1)},
\label{Phi-n-ansatz}
\end{equation}
 with $l$ to be determined. Note that values of $n$ here are not
 necessary
 integer but are to be found from the consistency with the resuling equations
 for $r_n(y)$.
 For the homogeneous equation (\ref{Phi-eq}), coefficients
$c_n$ are arbitrary. Each $r_n(y)$ is a solution of
 the equation
\begin{equation}
 y r_n^{\prime\prime} + L_n r_n^{\prime} + {\lambda\over (y+1)^2}\ r_n=0,
\label{rn-eq}
\end{equation}
 with an additional condition that it should be also consistent with
 the original integral equation (\ref{r-eq}).
 Here
\begin{equation}
 L_n=l+2-n.
\end{equation}

 The general solution is
\begin{widetext}
\begin{equation}
 r_n(y)= C_1 (y+1)^A F(a,b,c;y+1) +
    C_2 (y+1)^B F(a^\prime,b^\prime,c^\prime;y+1),
\label{rn-sol}
\end{equation}
 with
\begin{equation}
\begin{array}{ll}
 a=A,                    & a^\prime=B, \\
 b=L_n-1+A,              & b^\prime=L_n-1+B, \\
 c=2A,                   & c^\prime=2B, \\
 \displaystyle A={1\over 2} (1+ \sqrt{1+4\lambda} \, )
        =l+2 \ \ \
 &
 \displaystyle B={1\over 2} (1- \sqrt{1+4\lambda} \, )
        =1-A=-(l+1).
\end{array}
\label{rn-hg-sol}
\end{equation}
 This presentation has the form most suitable for the analysis
 of the consistency with the additional conditions.

 The Hypergeometric series in each of the terms of (\ref{rn-eq})
 becomes a finite sum, if $b$ ($b^\prime$) is a negative integer
 or zero; and the series does not converge for positive $y$
otherwize. For consistency with the integral equation, it is
necessary
 that all $r_n(y)$ vanish as $y\to\infty$. This leads to the
 following conditions:
 a) $C_1$=0,
 and b) $b^\prime$ should be a non-positive integer satisfying
 $-b^\prime<-B$.

 Combining all expressions, we obtain
\begin{equation}
 r_n(y)=   {C \over (y+1)^{l+1}}
      F\left (-(l+1),-n,-2(l+1);y+1 \right ), \ \ \ \
 n<l+1={1\over 2} \left (\sqrt{1+4\lambda} \, -1 \right ).
\label{rn-sol-final}
\end{equation}
 For $n=0$, this reproduces the solution Eq.(\ref{r0-solution}).
 For sufficiently small values of $\lambda$ ($\lambda<2$), it is
 the only solution. For $2<\lambda<6$, there two
 different solutions with $n=0$ and $n=1$, and so on.  We have
recovered the solutions already discussed in Section 3.

\end{widetext}

\section{}

In this Appendix it is shown that Eq.~(\ref{BSeqseries1}) can be approximated by the simpler Eq.~(\ref{BSeqseries2}).  This equation says that the leading $x_B$ dependence of $(n+1)$th order term is obtained from the $n$th order term by treating its $\kappa$ dependence as a constant, fixed by its value at $\kappa=0$.

It turns out that the best way to show this is to examine the structure of the first two terms in the series.   Dropping the unnecessary $Q^2$ dependence (and noting that the pole term does not depend on $\kappa$),  the first order term is
\bea
M^{(1)}(x_B,\kappa)&=&\lambda\int_{x_B}^1d\alpha \int_{\alpha\kappa}^\infty d\kappa' \frac{m^2\,M_{\rm pole}(x_B/\alpha)}{(m^2+\kappa')^2}  \nonumber\\
&=&C_0\int_{x_B}^1 d\alpha \left(\frac{m^2}{m^2+\alpha \kappa}\right)\frac{x_B^2}{\alpha^2}\,\delta\left(\frac{x_B}{\alpha} -1\right)\nonumber\\
&=&C_0\,x_B \frac{m^2}{m^2+x_B\kappa}\to C_0\,x_B\, ,
\eea
where the last is exact if $\kappa=0$.  Calculation of the second order term reveals the pattern that will allow us to calculate and sum the {\it leading small $x_B$ behavior\/} of all the terms.  Inserting the exact result for $M^{(1)}$ gives
\bea
M^{(2)}(x_B,\kappa)&=&\lambda C_0 x_B\int_{x_B}^1 \frac{d\alpha}{\alpha}\label{B2} \\
&&\times\int_{\alpha\kappa}^\infty d\kappa' \frac{m^2}{(m^2+\kappa')^2}
\left(\frac{m^2}{m^2+\frac{x_B}{\alpha}\kappa'}\right)  \, .
\nonumber
\eea
However, the $\kappa'$ integral is dominated by the region near its lower limit $\kappa'\simeq\alpha\kappa$, and in this region the factor arising from the $\kappa$ dependence of $M^{(1)}$ is unity (up to terms of order $x_B$)
\bea
\left(\frac{m^2}{m^2+\frac{x_B}{\alpha}\kappa'}\right) \simeq
\left(\frac{m^2}{m^2+x_B\kappa}\right) \to 1+{\cal O}(x_B)\, .
\label{B3}
\eea
Hence the leading order result for $M^{(2)}$ is simply
\bea
M^{(2)}(x_B,\kappa)\simeq\lambda C_0 x_B\int_{x_B}^1 \frac{d\alpha}{\alpha}
\int_{\alpha\kappa}^\infty d\kappa' \frac{m^2}{(m^2+\kappa')^2}    \, ,\qquad \label{M2}
\eea
precisely what would have been obtained if $M^{(1)}$ had been evaluated at $\kappa=0$ from the start, and its $\kappa$ dependence ignored.   Furthermore, the structure of Eq.~(\ref{M2}) shows that the leading $x_B$ dependence of $M^{(2)}$ is obtained by setting $\kappa=0$.   When $x_B\to0$, the $\alpha$ integral is dominated by values of $\alpha$ near $x_B$, and hence
\bea
\int_{\alpha\kappa}^\infty d\kappa' \frac{m^2}{(m^2+\kappa')^2}    &\to& \int_{x_B\kappa}^\infty d\kappa' \frac{m^2}{(m^2+\kappa')^2} \nonumber\\
\to \frac{m^2}{(m^2+x_B \kappa)}  &\to& 1 +{\cal O}(x_B)\, . \qquad\label{B5}
\eea
This confirms that the leading order contribution to $M^{(2)}$ is obtained from
\bea
M^{(2)}_{\rm leading}(x_B)=\lambda \int_{x_B}^1 d\alpha\;
M^{(1)}_{\rm leading}(x_B/\alpha)
\eea
in full agreement with Eq.~(\ref{BSeqseries2}).  This can be confirmed order by order, proving (\ref{BSeqseries2}).

We conclude by confirming by direct calculation some of our previous arguments.  First, the exact result for the $\kappa'$ integral in (\ref{B2}) is
\bea
I_{B2}=\int_{\alpha\kappa}^\infty d\kappa' \frac{m^2}{(m^2+\kappa')^2}
\left(\frac{m^2}{m^2+\frac{x_B}{\alpha}\kappa'}\right)
=\frac{\alpha} {\alpha-x_B}\quad
 \nonumber\\
\quad\times
\left\{\frac{m^2}{m^2+\kappa}+\frac{x_B} {\alpha-x_B} \log
\left[1-\frac{m^2(\alpha-x_B)} {\alpha(m^2+x_B\kappa)}\right]\right\}\, . \quad
\eea
This is regular, even when $\alpha\to x_B$, and hence
\bea
\lim_{x_B\to0} I_{B2}=\frac{m^2}{m^2+\alpha\kappa}\, ,
\eea
which is the result we obtain using the approximation ({\ref{B3}).

Next, we check the approximation used to obtain $M^{(2)}$ by evaluating the integral in (\ref{M2}) exactly:
\bea
\int_{x_B}^1 \frac{d\alpha}{\alpha}\frac{m^2}{m^2+\alpha\kappa}= \log\frac{1}{x_B}-\log\frac{m^2+\kappa}{m^2+x_B\kappa}\, .
\eea
When $x_B$ is small, only the $\log(1/x_B)$ term survives, which is  equal to the exact result at $\kappa=0$.  This confirms the correctness of the approximation (\ref{B5}).


\end{document}